\documentclass[12pt]{iopart}
\usepackage{graphicx}

\begin{document}
\letter{
A theory of new type of heavy-electron superconductivity 
in PrOs$_4$Sb$_{12}$: 
quadrupolar-fluctuation mediated odd-parity pairings
}

\author{K Miyake\dag, H Kohno\dag and H Harima\ddag}

\address{\dag\ Division of Materials Physics, Department of Physical Science, 
Graduate School of Engineering Science, 
Osaka University, Toyonaka, Osaka 560-8531, Japan
}
\address{\ddag\ 
The Institute of Scientific and Industrial Research, 
Osaka University, Ibaraki, Osaka 567-0047, Japan
}

\ead{miyake@mp.es.osaka-u.ac.jp}

\begin{abstract}
It is shown that unconventional nature of superconducting state of 
PrOs$_4$Sb$_{12}$, a Pr-based heavy electron compound with the 
filled-Skutterudite structure, can be explained in a unified way by taking 
into account the structure of the crystalline-electric-field (CEF) level, 
the shape of the Fermi surface determined by the band structure calculation, 
and a picture of the quasiparticles in f$^{2}$-configuration with 
magnetically singlet CEF ground state.  Possible types of pairing are 
narrowed down by consulting recent experimental results.  
In particular, the chiral ``p"-wave states such as $p_{x}+{\rm i}p_{y}$ is 
favoured under the magnetic field due to the orbital Zeeman effect, 
while the ``p"-wave states with two-fold symmetery such as $p_{x}$ can be 
stabilized by a feedback effect without the magnetic field.  
It is also discussed that the double superconducting transition 
without the magnetic field is possible due to 
the spin-orbit coupling of the ``triplet" Cooper pairs in the chiral state.  
\end{abstract}

\submitto{JPCM}

\maketitle

Recently, the superconductivity has been found in heavy electron compound 
PrOs$_4$Sb$_{12}$ with the crystal structure of the filled 
Skutterudite~\cite{Bauer}.  Since the specific heat jump $\Delta C$ at the 
superconducting transition temperature $T_{\rm c}$=1.8K is quite enhanced 
as $\Delta C/T_{\rm c}\simeq 500$ mJ/K$^2$mole, heavy quasiparticles are 
responsible for the Cooper pair formation.  Quite recently, a measurement 
of the longitudinal relaxation rate, $1/T_1$, of NQR at Sb site has been 
performed and very unusual temperature ($T$) dependence was revealed 
both for $T<T_{\rm c}$ and $T>T_{\rm c}$~\cite{Kotegawa}, 
while the normal state properties had been known to be also quite 
unconventional~\cite{Bauer,Aoki}.  Unconventional behaviors of 
$1/T_1$ are summarized as follows: 
1) Pseudo gap behavior is seen in $1/T_{1}T$ at $T_{\rm c}<T<2T_{\rm c}$, 
in which the resistivity $\rho$ also shows a pseudo-gap 
behavior~\cite{Bauer}.
2) There is no trace of the coherence (Hebel-Slichter) peak around 
$T=T_{\rm c}$ at all.  
3) $1/T_1$ appears to exhibit an exponential $T$-dependence below 
$1.3T_{\rm c}$, giving the superconducting gap $\Delta$ in the low 
temperature limit as $2\Delta/k_{\rm B}T_{\rm c}\simeq 5.3$, although 
a possibility is not ruled out that the crossover to the $T^{3}$-dependence 
begins to be observed at around $T\simeq T_{\rm c}/3$ the lowest 
temperature covered by experiments.  

Very recently, the anomaly of specific heat near $T_{\rm c}$ has been 
observed, which suggests a double transition at $T=T_{\rm c1}$ and 
$T_{\rm c2}$ ($T_{\rm c2}<T_{\rm c1}$)~\cite{Aoki,Vollmer}  
It also turned out very recently on the basis of measurements of the 
angular dependence of the thermal conductivity $\kappa$ under the magnetic 
field $H$ \cite{Izawa1} that there exsist at least two different 
superconducting phases in the $T$-$H$ phase diagram.  
In the low-field phase, the 2-fold component of $\kappa_{z}$ along the 
$z$-direction is observed as a function of the angle of the direction of 
$H$ around the $z$-axis, while the 4-fold one is observed in the high-field 
phase.  The phase boundary approaches 
the lower critical temperature $T_{\rm c2}$ as $H\to 0$.  This is 
in marked contrast with the case of a heavy electron superconductor 
CeCoIn$_5$ where the data can be interpreted by a simple 
``d"-wave model~\cite{Izawa2}.  
These behaviors suggest that a novel type of heavy electron 
superconductivity is realized in PrOs$_4$Sb$_{12}$.  

The purpose of this letter is to present a scenario explaining 
such anomalous behaviors in a unified way on the basis of the 
crystalline-electric-field (CEF) level, inferred from the experiment 
\cite{Bauer} and theoretical study \cite{Takegahara}, a 
topology of the Fermi surface (FS) offered from the band structure 
calculations \cite{Harima}, and a picture of the quasiparticles of 
$f^{2}$-based heavy electrons with a non-Kramers (non-magnetic) doublet 
CEF ground state.  

The CEF level scheme proposed by Bauer {\it et al.}~\cite{Bauer} 
in the point group $T_{h}$ is as follows~\cite{Takegahara}: 
The lowest level is the non-Kramers doublet $\Gamma_{23}^{\pm}$ 
($\Gamma_3^{\pm}$ in the representation of the point group $O_{h}$) 
\begin{eqnarray}
|\Gamma_{23}^{+}\rangle&=&\sqrt{\frac{7}{24}}(|+4\rangle+|-4\rangle)
-\sqrt{\frac{5}{12}}|0\rangle ,
\label{Gamma23+}
\\
|\Gamma_{23}^{-}\rangle&=&\frac{1}{\sqrt{2}}(|+2\rangle+|-2\rangle) ,
\label{Gamma23-}
\end{eqnarray}
and the first excited state is one of the triplet states $\Gamma_{4}^{(1)}$ 
or $\Gamma_{4}^{(2)}$  ($\Gamma_{4}$ or $\Gamma_{5}$ for $O_{h}$), wavefunction of which has the form 
\begin{equation}
|\Gamma_{4}^{(i)}\rangle=
\cases{
A_{1}^{(i)}(|-4\rangle-|4\rangle)+
A_{2}^{(i)}(|-2\rangle-|2\rangle)
 &
\label{Gamma40}\cr
\quad &\ \cr
B_{1}^{(i)}|\mp3\rangle+B_{2}^{(i)}|\mp1\rangle+
B_{3}^{(i)}|\pm1\rangle+B_{4}^{(i)}|\pm3\rangle ,
 &
\label{Gamma4pm}\cr}
\end{equation}
where the coefficients $A^{(i)}$'s and $B^{(i)}$'s are not 
universal but depend on the details of the CEF 
parameters~\cite{Takegahara}.  The lowest excitation energy of CEF levels 
has been estimated as $\Delta_{\rm CEF}\simeq 7$K~\cite{Vollmer}.  
Other excited CEF levels have excitation 
energies higher than 100K so that their effects 
are negligible in the low temperature region of the order of $T_{\rm c}$.  
However, the possibility of $\Gamma_{1}$-$\Gamma_{4}$ CEF level scheme, with 
$\Delta_{\rm CEF}\simeq 7$K, cannot be ruled out only from the analysis of the 
static susceptibility and the specific heat~\cite{Aoki}  
The ultrasonic measurements performed quite recently strongly suggest that the 
$\Gamma_{23}$ is the CEF gorund state~\cite{Goto}

Around $T\sim$10K, these lowest excited CEF levels 
give considerable contribution not only to the thermodynamic 
quantities, such as the specific heat $C$ and the magnetic susceptibility 
$\chi$, but also to the NQR relaxation rate $1/T_1$, since the 
``spin-flip" process can occur among the states forming $\Gamma_{4}^{(i)}$, 
e.g., between $|\pm1\rangle$ and $|\pm2\rangle$, giving the NQR 
relaxation.  It is noted that each CEF level broadens due to the 
hybridization with the conduction electrons so that the energy conservation 
law is satisfied in the NQR or NMR relaxation process.  
Indeed, if we assume that the energy level of excited CEF level is broadened 
such that its spectral weight is approximated by the Lorentzian with the 
width $\Delta E$, and that the processes across the CEF ground 
states, (\ref{Gamma23+}) and (\ref{Gamma23-}), and $\Gamma_{4}$'s are 
neglected, the imaginary part of the spin-flip susceptibility 
${\rm Im}\chi_{\perp}(\omega)$ is given simply, in the limit $|\omega|\ll T$, 
as 
\begin{equation}
{\rm Im}\chi_{\perp}(\omega)\simeq{\rm const.}\times{\omega\Delta E 
\over \pi T(\omega^{2}+\Delta E^{2})}e^{-\Delta_{\rm CEF}/T},
\label{chitrans}
\end{equation}
where const. is given by a combinatin of 
the coefficients $A$'s and $B$'s in (\ref{Gamma40}), and 
the Clebsch-Gordon coefficeints.  
Therefore, the NQR/NMR relaxation rate 
$1/T_{1}^{\rm CEF}\approx A_{\rm hf}^{2}T{\rm Im}\chi_{\perp}(\omega)/\omega$ 
due to the excited CEF level is given as 
\begin{equation}
{1\over T_{1}^{\rm CEF}}
\simeq{\rm const.}\times{1\over \pi\Delta E}e^{-\Delta_{\rm CEF}/T}
\label{rate}
\end{equation}
The width of CEF level arises from the hybridization between f- and conduction 
electrons and is of the order of the width of the renormalized quasiparticle 
band.  In the present case, $\Delta_{\rm CEF}\simeq 7$K is comparable to 
the bandwidth of heavy electrons, so that $\Delta E$ is also expected to be 
highly renormalized by the correlation effect.  

Namely, if the temperature is decreased well below $\Delta_{\rm CEF}$=7K, the 
relaxation processes are gradually killed, leading to the pseudo-gap behavior 
(\ref{rate}) at such temperature region.  We have, however, the 
usual relaxation process due to the quasiparticles of the Fermi liquid 
in addition.  
Therefore, $T$-dependence of $1/T_1=1/T_{1}^{\rm CEF}+1/T_{1}^{\rm qp}$ 
at $T<T_{\rm c}$ will be rather complicated one, 
since both contributions, from $\Gamma_4$ CEF level ($1/T_{1}^{\rm CEF}$) 
and the quasiparticles ($1/T_{1}^{\rm qp}$), 
to $1/T_1$ are decreasing in such $T$-region with different $T$-dependence 
in general.  In particular, one has to be careful in drawing the structure of 
the superconducting gap from the $T$-dependence of $1/T_1$ at 
$T_{\rm c}/4<T<T_{\rm c}$.

\begin{figure}
\begin{center}
\includegraphics[width=0.4\linewidth]{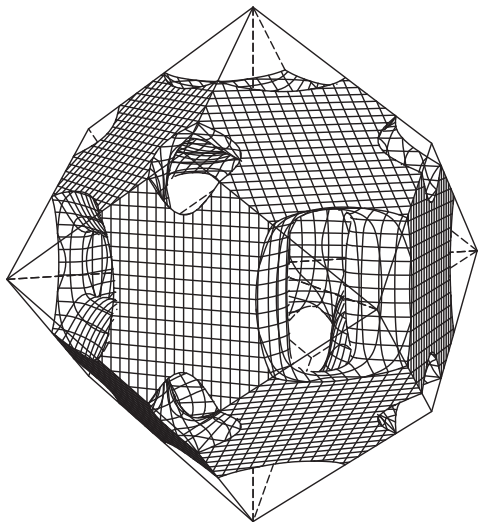}
\caption{
Fermi surface of PrOs$_4$Sb$_{12}$ relevant to the heavy electrons given by 
band structure calculation~\cite{Harima}.  FS is missing in the direction 
of [1,1,1] or [1,0,0], and its equivalent ones.}
\label{fig:1}
\end{center}
\end{figure}

In this Letter, we discuss the nature of the gap structure specific to 
the present 
system.  First of all, it may be reasonable to assume that the strong on-site 
repulsion, the possible origin of the heavy electron state, cannot be avoided 
in a manifold of the conventional s-wave pairing state.  This is also 
consistent with the absence of the coherence peak in $1/T_1$ ~\cite{Kotegawa} 
although the possibility of the strong coupling effect is not completely 
ruled out.  Another 
crucial aspect of PrOs$_4$Sb$_{12}$ uncovered by the band structure 
calculation is that the FS of the heavy electron band 
is missing in the directions of [1,0,0] and [1,1,1] and their equivalents 
as shown in Fig.\ \ref{fig:1}~\cite{Harima}.  There exists a small FS 
surrounding the $\Gamma$-point whose mass is not heavy~\cite{Harima} 
and has been detected by the de Haas-van Alphen experiment~\cite{Sugawara}.  
Due to this porous structure of FS, even the anisotropic 
pairing state can have finite gap over the FS.  However, even with such FS 
features being taken into account, the anisotropy of the gap due to such 
features of FS does not seem to fully explain the exponential-like 
$T$-dependence of $1/T_1$.  
\\
  \\
{\it Odd-parity pairing due to quadrupolar fluctuations}\par
\begin{figure}
\begin{center}
\includegraphics[width=0.3\linewidth]{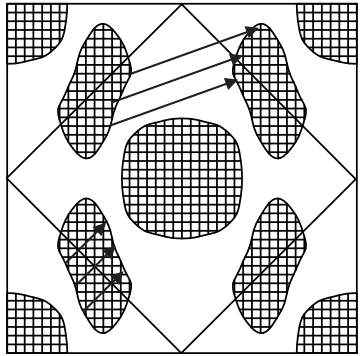}
\caption{
Fermi surface shown in Fig.\ \ref{fig:1} cut by the $k_{x}$-$k_{y}$ plane 
$k_{z}=0$. The nesting tendency in the heavy electron band 
at $(k_x,k_y)=(\pi/2,\pi/2)$ and 
$(3\pi/2,\pi/2)$ as shown by arrows, and their equivalent positions, 
remains in the direction rather rubustly for finite $k_{z}\not=0$. }
\label{fig:2}
\end{center}
\end{figure}

We adopt here the odd-parity pairing to explain the unconventional 
nature of the superconducting state mentioned above.  There are at least 
three circumstantial evidences for favouring the odd-parity pairing.  
First, the pairing interaction should be mainly mediated by the mode 
which gives rise to mass enhancement 
of quasiparticles, the quadrupolar fluctuations in the present 
case.  Quadrupolar susceptibility $\chi_{Q}({\bf q})$ is expected to 
be enhanced at large wave vector because the main FS has the 
nesting tendency as shown in Fig.\ \ref{fig:2} which can induce the 
attraction in both the ``d"- and ``p"-wave channel since the spin 
factor $({\vec \sigma}\cdot{\vec \sigma}^{\prime})$ does not exist in 
contrast to the case of spin-fluctuation mechanism~\cite{Nakajima,MSRV}.  
Second, a scenario for the bouble transition 
is more easily constructed in the odd-parity pairing with degeneracy 
due to the time-reversal symmetry than the even-parity pairings.  
Third, the so-called Maki parameter $\kappa_{2}$ under the 
magnetic field exhibits no paramagnetic limitation~\cite{Aoki2}.  

As mentioned above, the pairing is also expected to be induced by exchanging 
the quadrupolar fluctuations of the non-Kramers doublet 
$\Gamma_{23}^{\pm}$.  
The quadrupolar coupling between essentially 
localized 4f$^{2}$-states and the quasiparticles containing considerable 
weight of the conduction electrons, arises through the hybridisation with the 
local symmetry of $\Gamma_{8}^{(1)}$ and $\Gamma_{8}^{(2)}$ in the cubic 
representaion of CEF state of $j=5/2$-manifold, as discussed by 
Cox in the context of the quadrupolar Kondo effect~\cite {Cox}.  The fact that 
the heavy quasiparticles contains the considerable weight of conduction 
electrons is a salient feature of f$^{2}$-based heavy electron state, which 
is in marked contrast with the f$^{1}$- or f$^{3}$-based ones where 
the quasiparticles are dominated by f-electrons.  It is also consistent 
with the result of band structure calculation which shows that 
the f-component of the heaviest band at FS is only several \%~\cite{Harima}.  
The propagator of the quadrupolar fluctuations 
$\chi_{Q}(q,{\rm i}\omega_{m})$ may be given as 
\begin{equation}
\chi_{Q}(q,{\rm i}\omega_{m})={\chi_{Q}^{\ell{\rm oc}}({\rm i}\omega_{m})\over 
1-g^{2}\Pi_{0}(q,{\rm i}\omega_{m})},
\label{chiquadrupolar}
\end{equation}
where $\chi_{Q}^{\ell{\rm oc}}$ and $\Pi_{0}$ denote the propagator for local 
fluctuations of quadrupolar moment, and the polarisation function of 
quasiparticles, respectively, and $g$ is the coupling constant among them.  

As shown in Fig.\ \ref{fig:2}, the FS has a nesting tendency and 
$\Pi_{0}$ is expected to have peaks at 
${\vec q}=(\pi/2,\pi/2,0)$, and ${\vec q}=(\pi/2,3\pi/2,0)$, and their 
equivalent positions.  It is noted that the FS is rather flat in the 
$z$-direction near the nesting position as can be seen in Fig.\ \ref{fig:1}.  
Then, the pairing interaction in the static approximation, 
$\Gamma({\vec q})\simeq g^{2}\chi_{Q}(q,0)$, can be parameterised as 
\begin{eqnarray}
\Gamma({\vec q})&=&\Gamma_{0}-\Gamma_{1}[\cos(2q_{x})+\cos(2q_{y})]
+\Gamma_{2}\cos{q_{x}\over 2}\cos{q_{y}\over 2}
\nonumber \\
& &\quad
+({\hbox{cyclic permutations of $q_{x}$, $q_{y}$, and $q_{z}$}}),
\label{Gamma1}
\end{eqnarray}
where $\Gamma_{i}$'s are positive constants and $\Gamma_{2}$ is rather 
smaller than than $\Gamma_{1}$.  The term of $\Gamma_{2}$ represents 
the effect that tendency of the nesting at ${\vec q}=(\pi/2,3\pi/2,0)$ is 
less than that at ${\vec q}=(\pi/2,\pi/2,0)$ as seen in Fig.\ \ref{fig:2}.  
By putting ${\vec q}={\vec k}-{\vec k}^{\prime}$, 
$\Gamma({\vec q})$ is represented near the peak as follows:
\begin{equation}
\Gamma({\vec k}-{\vec k}^{\prime})=\Gamma_{0}
-\Gamma_{1}[\cos2(k_{x}-k_{x}^{\prime})+\cos2(k_{y}-k_{y}^{\prime})]
+\cdots
\label{Gamma2}
\end{equation}
This gives the attractive interactions in the following channels: 
\begin{eqnarray}
{\hbox{``d"-wave;\ }}
& &\cos(2k_{x})-\cos(2k_{y}),{\hbox{\ etc.}},
\label{Gamma3} \\
{\hbox{``p"-wave;\ }}
& &\sin(2k_{x}),\quad \sin(2k_{y}),\quad \sin 2k_{z},
\label{Gamma4}
\end{eqnarray}
Among these states, $\sin(2k_{x})$ and its equivalents, 
$\sin(2k_{y})$ and $\sin(2k_{z})$, will be the most favourable ones 
because they have maximum amplitude on the FS.  Indeed, other states 
have more nodes on the FS as seen in Fig.\ \ref{fig:2}.  

The simplest odd-parity states with ``equal-spin-pairing" (ESP) allowed 
in cubic symmetry are given as follows:
\begin{eqnarray}
\hat{\Delta}_{k}&=&\Delta[p_{x}(k)+\varepsilon p_{y}(k)
+\varepsilon^{2}p_{z}(k)]{\rm i}(\sigma_{y}\sigma_{x}),
\label{triplet1} \\
\hat{\Delta}_{k}&=&\Delta[p_{x}(k)+{\rm i}p_{y}(k)]
{\rm i}(\sigma_{y}\sigma_{x}),
\label{triplet2}
\end{eqnarray}

Here, 
${\sigma}_{j}$ is the $j$-th component of the Pauli matrix, 
$\varepsilon\equiv e^{{\rm i}2\pi/3}$, and $p(k)$'s are bases of irreducible 
representations with ``p"-symmetry: 
$p_{x}(k)\equiv\sqrt{2}\sin(2k_{x})$, etc..  
These gaps vanish along the direction 
of [1,1,1] or [1,0,0], and its equivalent direction.  
However, since there exists no FS in those directions, 
$1/T_1$ exhibits an exponential $T$-dependence in the lowest temperature 
region in spite of the anisotropic gap.  
It is noted that the Fermi surface of light electrons, detected by the 
de Haas-van Alphen experiment~\cite{Sugawara}, is closed surrounding 
the $\Gamma$-point and these gaps have nodes at points on this FS.  
However, since $1/T_1$ is proportional to the 
square of the density of states at the Fermi level, the effect of 
such light electrons should hardly be seen by the $T$-dependence of $1/T_1$.  

Other possible states in the odd-parity manifold are 
\begin{eqnarray}
\hat{\Delta}_{k}&=&\Delta p_{x}(k){\rm i}(\sigma_{y}\sigma_{x}),\quad
\hbox{{\rm and its equivalent ones}}.
\label{triplet3}
\end{eqnarray}
Such states are less favourable 
compared to the chiral states (\ref{triplet1}) and (\ref{triplet2}) in the 
so-called weak-coupling case where the feedback effect is not taken into 
account.  This can be seen from the structure of the GL free energy.  
For instance, in the case of odd-parity class of ESP with 
``p"-symmetry, it is given as follows~\cite{Leggett,Miyake}:
\begin{eqnarray}
& &F_{\rm GL}(\Delta_x,\Delta_y,\Delta_z)=F_{0}+
\Phi(1-V\Phi)(|\Delta_x|^{2}+|\Delta_y|^{2}+|\Delta_z|^{2})\qquad
\nonumber \\
&+&{\textstyle{1\over 2}}\chi_{\rm diag}
(|\Delta_{x}|^{4}+|\Delta_{y}|^{4}+|\Delta_{z}|^{4})
\nonumber \\
&+&\chi_{\rm off}
\biggl\{|\Delta_{x}|^{2}|\Delta_{y}|^{2}+|\Delta_{y}|^{2}|\Delta_{z}|^{2}
+|\Delta_{z}|^{2}|\Delta_{x}|^{2}
\nonumber \\
&+&2[{\rm Re}(\Delta_{x}\Delta_{y}^{*})]^{2}+
2[{\rm Re}(\Delta_{x}\Delta_{y}^{*})]^{2}+
2[{\rm Re}(\Delta_{x}\Delta_{y}^{*})]^{2}\biggr\},
\label{GL1}
\end{eqnarray}
where $\Delta$'s are the coefficient of each (normalized) irreducible 
representations of the gap, $V$ is the strength of pairing interaction of 
``p"-symmetry, and 
\begin{eqnarray}
\Phi&\equiv&
\sum_{k}[p_{x}(k)]^{2}{\tanh(\xi_{k}/2T)\over 2\xi_{k}},
\label{GL2} \\
\chi_{\rm diag}&\equiv&
\sum_{k}[p_{x}(k)]^{4}\left(-{{\rm d}\over {\rm d}\xi_{k}^{2}}
{\tanh(\xi_{k}/2T)\over 2\xi_{k}}\right)>0,
\label{GL3} \\
\chi_{\rm off}&\equiv&
\sum_{k}[p_{x}(k)p_{y}(k)]^{2}\left(-{{\rm d}\over {\rm d}\xi_{k}^{2}}
{\tanh(\xi_{k}/2T)\over 2\xi_{k}}\right)>0. 
\label{GL4}
k\end{eqnarray}
Since $\chi_{\rm diag}\ge\chi_{\rm off}$ due to the Schwarz inequality, 
the gap (\ref{triplet3}) cannot minimize $F_{\rm GL}$ in general.  
In case $\chi_{\rm off}<{1\over 3}\chi_{\rm diag}$, the gap 
(\ref{triplet1}) minimizes $F_{\rm GL}$, while 
the gap (\ref{triplet2}) minimizes $F_{\rm GL}$ in case 
$\chi_{\rm diag}>\chi_{\rm off}>{1\over 3}\chi_{\rm diag}$.  

The low-field phase, in the $T$-$H$ phase diagram~\cite{Izawa1}, 
having 2-fold symmetry is consistent 
with the gap (\ref{triplet3}), which cannot be 
realized in the weak-coupling theory.  This is also the case in 
the singlet manifold.  The so-called BW-like state is known to be 
the most stable state in the weak-coupling approximation, and there 
is no reason in principle to rule out its possibility from the 
first.  However, BW-like state looks inconsistent with the 
thermal conductivity measurement under the magnetic field~\cite{Izawa1} 
and other thermodynamic measurements.  
\\
  \\
{\it Feedback effect}\par
In order that the gap with 2-fold symmetry such as (\ref{triplet3}) 
to be realised, we need a feedback effect.  Among them, the 
following mechanism may be promising.  
The polarisation function $\Pi_{0}$ appearing (\ref{chiquadrupolar})  
in the superconducting state is given as 
\begin{equation}
\Pi_{0}(q,0)={1\over 2}\sum_{{\vec k}}
{E_{{\vec k}}E_{{\vec k}+{\vec q}}-\xi_{{\vec k}}\xi_{{\vec k}+{\vec q}}
+\Delta_{{\vec k}}\Delta_{{\vec k}+{\vec q}}\over 
E_{{\vec k}}E_{{\vec k}+{\vec q}}(E_{{\vec k}}+E_{{\vec k}+{\vec q}})},
\label{Pi}
\end{equation}
where $E_{{\vec k}}=\sqrt{\xi_{{\vec k}}^{2}+|\Delta_{{\vec k}}|^{2}}$.  
If the nesting were perfect at 
${\vec q}=(\pi/2,\pi/2,0)$, and ${\vec q}=(\pi/2,3\pi/2,0)$, and their 
equivalent positions, the following relations would hold, 
$\xi_{{\vec k}+{\vec q}}=-\xi_{{\vec k}}$, 
$\Delta_{{\vec k}+{\vec q}}=-\Delta_{{\vec k}}$, and 
$E_{{\vec k}+{\vec q}}=E_{{\vec k}}$, for ${\vec k}$ near the FS.  
Then, the expression, (\ref{Pi}), would be reduced to 
\begin{equation}
\Pi_{0}(q,0)={1\over 2}\sum_{{\vec k}}
{\xi_{{\vec k}}^{2}\over 
(\xi_{{\vec k}}^{2}+|\Delta_{{\vec k}}|^{2})^{3/2}}
\label{Pi2}
\end{equation}
Then, the polarisation mediating the pairing interaction depends on the 
type of pairing itself.  Indeed, the pairing (\ref{triplet3}) is 
expected to give larger (\ref{Pi2}) than the pairing (\ref{triplet2}), because 
the gap function of (\ref{triplet3}) 
\begin{equation}
|\Delta_{{\vec k}}|^{2}\propto2\sin^{2}(2k_{x})
\label{Pi3}
\end{equation}
vanishes on the planes, $k_{x}=0,\pm\pi/2,\pm\pi$, and $\pm3\pi/2$, 
which pass through near the FS, while the gap function (\ref{triplet2}) 
\begin{equation}
|\Delta_{{\vec k}}|^{2}\propto[\sin^{2}(2k_{x})+\sin^{2}(2k_{y})]
\label{Pi4}
\end{equation}
vanishes only on the lines, $(k_{x},k_{y})=(\pm\pi/2,\pm\pi/2)$, 
$(\pm\pi/2,\pm\pi)$, $(\pm\pi/2,\pm3\pi/2)$, etc., which are located away 
from the FS.  Although the explicit band structure calculation is hard 
in practice for the moment, the tendency mentioned above 
is expected to remain valid.  Therefore, the state (\ref{triplet3}) 
may be stabilized against (\ref{triplet2}) by the feedback effect.  
In the BW-like state, $\Pi_{0}(q,0)$, (\ref{Pi2}), is suppressed 
more severely than in the state (\ref{triplet2}), and is destabilized 
against (\ref{triplet3}).  

This kind of feedback effect is an analogue of that due to the 
ferromagnetic spin-fluctuation mechanism discussed in superfluid 
$^3$He~\cite{AB,Kuroda}, in which the spin-fluctuation spectrum 
depends on the gap structure of the triplet states.  \\
  \\
{\it Double transition due to spin-orbit coupling}\par
The spin-orbit interaction $H_{\rm so}$ due to the mutual Coulomb interaction 
between electrons and relative motion is given by 
\begin{equation}
H_{\rm so}=-{\mu_{\rm B}^{2}\over 2\hbar}{m\over m_{\rm band}}
\sum_{i}\sum_{j\not=i}
{1\over r_{ij}^{3}}({\vec{\sigma}}_{i}+{\vec{\sigma}}_{j})\cdot
\left[{\vec r}_{ij}\times[(2{\bar g}-1){\vec p}_{i}
-2{\bar g}{\vec p}_{j}]\right],
\label{spinorbit1}
\end{equation}
where $\mu_{\rm B}$ is the Bohr magneton, $m$ electron mass, $m_{\rm band}$
the band mass, and ${\bar g}$ is defined 
as ${\bar g}\equiv\mu_{\rm eff}/\mu_{\rm B}$, $\mu_{\rm eff}$ being 
the effective magnetic moment 
$\mu_{\rm eff}\equiv(6/7)|\langle j_{z}\rangle|\mu_{\rm B}$.  
The appearance of the factor $m/m_{\rm band}$ in (\ref{spinorbit1}) 
can follows from the Ward-Pitaevskii-identity~\cite{Miyake2}.  
By the procedure similar to that described in Ref.\ \cite{Leggett} for 
the dipole interaction, the interaction (\ref{spinorbit1}) leads to 
the spin-orbit free energy $F_{\rm so}$ for Cooper pairs which is spin 
triplet and chiral, such as (\ref{triplet2}), 
with the pair angular momentum 
$\hbar{\vec{\ell}}$ as follows\cite{Miyake2}:
\begin{equation}
F_{\rm so}=-g_{\rm so}({\rm i}{\vec d}\times{\vec d}^{*})\cdot{\vec {\ell}},
\label{spinorbit2}
\end{equation}
where 
\begin{equation}
g_{\rm so}=g_{\rm D}{m\over m_{\rm band}}\times {20\over 3}(4{\bar g}-1)
=g_{\rm D}{m\over m_{\rm band}}\times 
\cases{{20\over 3}\times{37\over7},&for $\Gamma_{8}^{(2)}$;\cr
\quad&\ \cr
{20\over 3}\times{5\over 7},&for $\Gamma_{8}^{(1)}$,\cr}
\label{spinorbit3}
\end{equation}
where $g_{\rm D}$ is the strength of the dipole coupling in the 
``ESP"-superconducting state, and we have used 
$\langle j_{z}\rangle=\pm11/6$ for quasiparticles consisting of 
$\Gamma_{8}^{(2)}$ f$^{1}$-CEF state, and $\pm1/2$ for $\Gamma_{8}^{(1)}$.  
The free energy due to the dipole-dipole interaction is given as \cite{Leggett}
\begin{equation}
F_{\rm D}=-{3\over 5}g_{\rm D}|({\vec d}\cdot{\vec {\ell}})|^{2}.
\label{dipole}
\end{equation}
Therefore the spin-orbit interaction, in the non-unitary state with 
$|{\vec d}\times{\vec d}^{*}|\simeq 1$, dominates the dipole-dipole 
interaction, in the unitary state with $|{\vec d}\cdot{\vec{\ell}}|=1$, 
because $g_{\rm so}$ far exceeds $g_{\rm D}$ in $\Gamma_{8}^{(2)}$-band 
considering (\ref{spinorbit3}) and $m/m_{\rm band}\sim{\cal O}(10^{-1})$.  
Following the calculation in the 
spherical model~\cite{Leggett}, $g_{\rm D}$ is given by 
\begin{equation}
g_{\rm D}={F_{\rm cond}\over 1-T/T_{\rm c}}\times 3.1\mu_{\rm eff}^{2}N_{\rm F}
\left[\ln(1.14\epsilon_{\rm c}/k_{\rm B}T_{\rm c})\right]^{2},
\label{spinorbit4}
\end{equation}
where $N_{\rm F}$ is the density of states (DOS) of the quasiparticles, and 
$F_{\rm cond}$ is the condensation free energy 
\begin{equation}
F_{\rm cond}=-N_{\rm F}{4(\pi k_{\rm B}T_{\rm c})^{2}
\over7\zeta(3)\kappa}\left(1-{T\over T_{\rm c}}\right)^{2},
\label{cond}
\end{equation}
where $\kappa$ is the average of square of the magnitude of normalized gap 
function over the FS.  
The second factor in (\ref{spinorbit4}) is estimated as 
\begin{equation}
3.1\mu_{\rm eff}^{2}N_{\rm F}
\left[\ln(1.14\epsilon_{\rm c}/k_{\rm B}T_{\rm c})\right]^{2}
\simeq \left({\mu_{\rm eff}\over\mu_{\rm B}}\right)^{2}
\times1.4\times 10^{-3},
\label{spinorbit5}
\end{equation}
where we have assumed that the renormalized Fermi 
energy is $\epsilon^{*}_{\rm F}\simeq 10^{4}/300$ K, the number density of 
quasiparticles 
$N/V=2/(2/\sqrt{3}r_{\rm Pr-Pr})^{3}$, $r_{\rm Pr-Pr}$ being the distance 
between two nearest Pr ions.  Therefore, the spin-orbit coupling $g_{\rm so}$, 
(\ref{spinorbit3}), is estimated as 
\begin{equation}
g_{\rm so}={F_{\rm cond}\over 1-T/T_{\rm c}}{m\over m_{\rm band}}\times 
\cases{{20\over 3}\times{37\over 7}\times 1.4\times 10^{-3}
=4.9\times 10^{-1},&for $\Gamma_{8}^{(2)}$;\cr
\quad&\ \cr
{20\over 3}\times{5\over 7}\times1.4\times 10^{-3}
=6.6\times 10^{-2}&for $\Gamma_{8}^{(1)}$,\cr}.
\label{spinorbit6}
\end{equation}
The free-energy difference between (\ref{triplet3}) 
and (\ref{triplet2}), its non-unitary version with 
$|{\vec d}\times{\vec d}^{*}|\not=0$, is of the order of 10\% of 
$F_{\rm cond}$ in general, and $m/m_{\rm band}\sim{\cal O}(10^{-1})$ 
according to the band structure calculation~\cite{Harima}.  
Therefore, if the stable state is (\ref{triplet3}) due to 
the feedback effect as discused above, there occurs a double 
transition with splitting of the transition temperature being 
($T_{\rm c1}-T_{\rm c2})/T_{\rm c1}\simeq {\rm several\ \%}$, because 
the state (\ref{triplet2}) of non-unitary version is stabilized, due to 
the spin-orbit interaction (\ref{spinorbit2}), against (\ref{triplet3}) 
which is real state and has no spin-orbit coupling 
such as (\ref{spinorbit2}).  This width of 
splitting of double transition is consistent with the experimental 
observations~\cite{Aoki,Vollmer}.  Especially, the results by Aoki 
{\it et al}~\cite{Aoki}, suggesting the double transition remains 
rather robustly under the magnetic field, can be explained by the present 
mechanism.  

The chiral state (\ref{triplet1}) has also intrinsic magnetic moment 
along (1,1,1) direction, so that it can also give rise to the double 
transition as above.  However, this state gives the angualr dependence 
of the thermal conductivity opposite to the observation in the high-field 
phase ~\cite{Izawa1}  although the 4-fold behaviour is expected.  
\\
  \\
{\it Two phases in T-H phase diagram}\par
Finally, the multiphase diagram in $T$-$H$ plane determined by the thermal 
conductivity measurements under the magnetic field~\cite{Izawa1} may be understood as follows:  A crucial point is that the triplet state 
(\ref{triplet2}) has the intrinsic magnetic moment ${\vec M}_{\rm in}$ 
associated with the intrinsic angular momentum ${\vec L}_{\rm in}$ as 
${\vec M}_{\rm in}=\mu_{\rm B}(m/m_{\rm band}){\vec L}_{\rm in}/\hbar$, where 
$m^{*}$ is the effective mass of heavy quasiparticles, and 
${\vec L}_{\rm in}$ is given as 
\begin{equation}
{\vec L}_{\rm in}={N_{\rm in}\over 2}\hbar{\vec{\ell}},
\label{intrinsic}
\end{equation}
where $N_{\rm in}$ is the order of the superfluid electron 
density $N_{\rm s}$~\cite{Ishikawa,Miyake3,Kita}, while the lively 
disputes were performed concerning the size of $N_{\rm in}$, whether 
$N_{\rm in}\sim {\cal O}(N_{\rm s})$ or ${\cal O}(N_{\rm s}\cdot 
(T_{\rm c}/\epsilon_{\rm F}^{*})^{n})$ with $n=$ 1 or 2, 
about a quarter of century ago in the context of superfluid 
$^{3}$He~\cite{WV}.  
Very recently, the reality of this intrinsic magnetic moment has caused 
a renewed interest in the magnetic property of the chiral superconducting 
state of Sr$_2$RuO$_4$~\cite{Ishida}.  
At low enough temperature $T\ll T_{\rm c}$, the intrinsic magnetic 
moment is ${\vec M}_{\rm in}\simeq 
(N/2)\mu_{\rm B}(m/m_{\rm band}){\vec{\ell}}$.  
Therefore, the state (\ref{triplet2}) is stabilized under the magnetic field 
$H$ over the state (\ref{triplet3}), which has no intrinsic magnetic 
moment.  The transition between the two states occurs when 
\begin{equation}
N_{\rm F}(k_{\rm B}T_{\rm c})^{2}\times 10^{-1}
\sim N{m\over m_{\rm band}}\mu_{\rm B}H\left(1-{N_{\rm d}\over 4\pi}\right),
\label{boundary}
\end{equation}
where the left-hand side represents the difference of 
the condensation energy between the two superconducting phases at 
$T\ll T_{\rm c}$, and 
the right-hand side the energy gain in the chiral state (\ref{triplet2}) 
with the intrinsic angular mementum in the magnetic field $H$.  
We have included in (\ref{boundary}) the so-called the demagnetisation 
factor $N_{\rm d}$ which depends on the sample shape.  
In (\ref{boundary}), the energy due to the magnetic field arising from 
the intrinsic magnetisation itself is neglected because it is much smaller 
than the external field $\simeq H$ in question.  
The magnetic field giving the phase boundary beween low- and high-field 
phases, determined by the thermal conductivity, roughly agrees with the 
present estimation is in the same order 
as given by (\ref{boundary}), because $N_{\rm F}\sim N/\epsilon^{*}_{\rm F}$, 
$k_{\rm B}T_{c}/\epsilon_{\rm F}^{*}\sim 10\times(m/m^{*})$, 
$m^{*}/m_{\rm band}\sim10$, and $N_{\rm d}\sim 10^{-1}$.  A crucial 
prediction of the present scenario is that the phase boundary in the $T$-$H$ 
plane is dependent on the sample shape through the demagnetisation 
factor $N_{\rm d}$.  

The angular dependence of the thermal conductivity $\kappa_{z}$ reported 
in Ref.~\cite{Izawa1} may also be explained by the present scenario.  
If the state (\ref{triplet3}), $\Delta_{k}\propto p_{x}(k)$, is realised 
in the low-field phase due to the boundary effect, which works to 
align the extention of pair wavefunction, $\kappa_{z}$ takes maximum 
(minimum), when the magnetic field ${\vec B}$ is 
${\vec B}\parallel {\hat x}$ (${\vec B}\parallel {\hat y}$), 
in consistent with Ref.~\cite{Izawa1}.  
This can be understood by applying the argumet similar to 
Ref.~\cite{Izawa2}.  
If a type of the state (\ref{triplet2}) is realized 
in the high-field phase, the free energy takes minimum when 
the quantisation axis of intrinsic angular momentum is parallel to 
${\vec B}$ for which $\kappa_{z}$ is minimum~\cite{Izawa2}.  Therefore, when 
the direction of ${\vec B}$ is rotated by angle $\phi$ from the $x$-axis 
in the plane perpendicular to $z$-axis, $\kappa_{z}$ increases and 
reaches maximum at $\phi=45^{\circ}$ above which the stable quantisation 
axis changes from $x$- to $y$-axis, and then $\kappa_{z}$ decreases up to 
$\phi=90^{\circ}$ where the stable configuration is reached again.  Namely, 
the 2-fold dependence of $\kappa_{z}(\phi)$ taking maximum at 
$\phi=0^{\circ}$ can be possible in the low-field phase, 
and the 4-fold one taking minimum at $\phi=0^{\circ}$ and $\phi=90^{\circ}$ 
in the high-field phase~\cite{Izawa1}.  
\\

In conclusion, it is remarked that the magnetic susceptibility, 
both longitudinal and transverse, can be enhanced by electron correlations 
even if the mass enhancement arises from the degeneracy due to the non-Kramers 
doublet, i.e., electric quadrupolar moment, provided that there exists a 
perturbation which breaks the particle-hole symmetry, such as the repulsion 
among conduction electrons, as shown by the numerical renormalization group 
calculations for the impurity model~\cite{Kusu}.  
This is in marked contrast with the case of heavy electrons based on 
f$^{2}$-configuration with the singlet CEF ground state~\cite{Yotsu1}, 
where the static susceptibility along the easy axis due to quasiparticles 
is not enhanced while the NMR/NQR relaxation rates given by the dynamical 
transverse susceptibility is enhanced in proportion to a square of the 
mass-enhancement factor as observed in UPt$_3$~\cite{Tou}.  \\

We have benefited from informative conversations 
with Y. Aoki, T. Goto, K. Izawa, Y. Kitaoka, H. Kotegawa, Y. Matsuda, 
H. Sato, H. Sugawara.  
This work was supported by the Grant-in-Aid for COE Research Program 
(No. 10CE2004) by the Ministry of Education, Culture, Sports, Science and 
Technology.

\end{document}